\documentclass[%
aip,
amsmath,amssymb,
reprint,%
]{revtex4-1}
\usepackage{graphicx}
\usepackage{dcolumn}
\usepackage{bm}
\usepackage{upgreek}
\usepackage[colorlinks=true,linkcolor=blue,citecolor=blue]{hyperref}%
\hypersetup{allcolors = blue}

\usepackage[utf8]{inputenc}
\usepackage[T1]{fontenc}
\usepackage{tikz}
\newcommand\copyrighttext{%
	\footnotesize This article may be downloaded for personal use only. Any other use requires prior permission of the author and AIP Publishing. This article appeared in Appl. Phys. Lett. \textbf{118}, 162402 (2021) and may be found at \href{https://doi.org/10.1063/5.0048825}{https://doi.org/10.1063/5.0048825}.}
\newcommand\copyrightnotice{%
	\begin{tikzpicture}[remember picture,overlay]
	\node[anchor=south,yshift=10pt] at (current page.south) {\fbox{\parbox{\dimexpr\textwidth-\fboxsep-\fboxrule\relax}{\copyrighttext}}};
	\end{tikzpicture}%
}

\newcommand{\mr}[1]{\mathrm{#1}}  

\begin{document}
\copyrightnotice
\preprint{AIP/123-QED}
\title{Bistable nanomagnet as programmable phase inverter for spin waves}

\author{Korbinian Baumgaertl}
\affiliation{Laboratory of Nanoscale Magnetic Materials and Magnonics, Institute of Materials (IMX), \'Ecole Polytechnique F\'ed\'erale de
	Lausanne (EPFL), 1015 Lausanne, Switzerland}

\author{Dirk Grundler}
\email{dirk.grundler@epfl.ch}

\affiliation{Laboratory of Nanoscale Magnetic Materials and Magnonics, Institute of Materials (IMX), \'Ecole Polytechnique F\'ed\'erale de
	Lausanne (EPFL), 1015 Lausanne, Switzerland}
\affiliation{Institute of Microengineering (IMT), \'Ecole Polytechnique F\'ed\'erale de Lausanne (EPFL), 1015 Lausanne, Switzerland}

\date{\today}

\begin{abstract}
	 To realize spin wave logic gates programmable phase inverters are essential.  We image with phase-resolved Brillouin light scattering
microscopy propagating spin waves in a one-dimensional magnonic crystal consisting of dipolarly coupled magnetic nanostripes. We demonstrate
phase shifts upon a single nanostripe of opposed magnetization. Using micromagnetic simulations we model our experimental finding in a
wide parameter space of bias fields and wave vectors. We find that low-loss phase inversion is achieved, when the internal field of the
oppositely magnetized nanostripe is tuned such that the latter supports a resonant standing spin wave mode with odd quantization number at the given frequency. Our results are key for the realization of phase inverters with optimized signal transmission.
\end{abstract}

\pacs{}

\maketitle
Spin wave (SW) computing is promising for future low-power consuming data processing \cite{Chumak2015_MagnonLogic, Csaba20171471_SW_LOGICS,
doi:10.1063/5.0019328}.
A common approach for SW logic gates relies on encoding the logic output in the combined amplitude of two (or more) interfering SWs
\cite{DW1, doi:10.1063/1.2089147,doi:10.1063/1.2975235, doi:10.1063/1.2834714,PMID:25975283, PhysRevApplied.9.014033}. By inverting the phase
of one of the incoming SWs, the output level can be switched from constructive interference with high amplitude (logic '1') to destructive
interference with low amplitude (logic '0'). For technological applications an ideal phase inverter should be efficiently gateable, introduce
little SW attenuation and operate at the nanoscale.
In initial works phase inversion was achieved by exposing SWs to an inhomogeneous magnetic field created by a current carrying wire
\cite{doi:10.1063/1.2089147,doi:10.1063/1.2975235, doi:10.1063/1.2834714, PMID:25975283}. To obtain higher efficiency, voltage controlled
anisotropy \cite{PhysRevApplied.9.014033, Rana2019} and magnetic defects \cite{doi:10.1063/1.4953395,Baumgaertl2018,
doi:10.1021/acsami.9b02717} have been explored. In previous works such as Refs. \cite{Baumgaertl2018,doi:10.1021/acsami.9b02717} phase shift
were detected electrically by propagating spin-wave spectroscopy. Spatially resolved data were not provided. The critical dimension for the
phase inversion process and its optimization remained unclear.\\
In this work we use phase-resolved Brillouin light scattering microscopy ($\upmu\mr{BLS}$) \cite{doi:10.1063/1.2335627,Demidov2015,
BLS_wavevector} to spatially resolve SW wavefronts in a 1D MC with a programmable magnetic defect. We evidence that the previously reported
phase shift\cite{Baumgaertl2018} $\Delta \Theta$ occurs locally within the individual magnetic defect. In our experiment, its width amounts to
325~nm much smaller than the SW wavelength $\lambda$. So far experimentally observed phase shifts were concomitant with a reduction $\eta$ in
the transmitted SW amplitudes \cite{Baumgaertl2018,doi:10.1021/acsami.9b02717}, hindering the performance of the phase inverter. Using
micromagnetic simulations we show that the reduction in amplitude can be circumvented by tuning the eigenfrequency of the magnetic defect such
that resonant coupling is achieved. Our findings are promising for the realization of a low-loss nanoscale phase inverter in magnonics.\\
\begin{figure*}[tbh!]
	\centering
	\includegraphics[width=\textwidth]{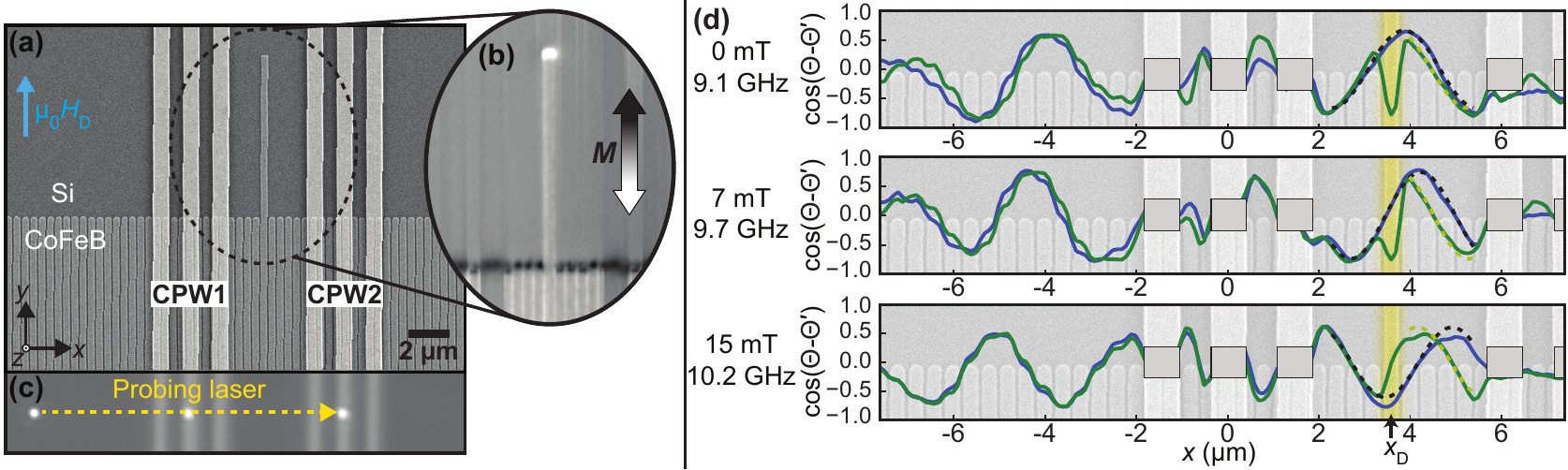}
	\caption{(a) SEM and (b) magnetic force microscopy image of the centeral region of MC1. The elongated nanostripe was intentionally
magnetized in $-y$-direction opposed to the magnetization of the short stripes in order to form a magnetic defect. (c) Optical image of MC1 as
seen in the BLS microscope. The probing laser spot was scanned along the yellow dashed line, while SW intensity and phase signal was recorded.
Microwave excitation was applied to CPW1. (d) Phase signal measured with (green lines) and without (blue lines) magnetic defect for different
$H_\mr{B}$ (rows). The position $x_\mr{D}$ of the magnetic defect is highlighted.}
	\label{fig:BLS_MC}
\end{figure*}
Figure \ref{fig:BLS_MC}(a) shows a scanning electron microscopy (SEM) image of the investigated sample. The 1D MC consisted of dipolarly
coupled $\mathrm{Co_{20}Fe_{60}B_{20}}$ nanostripes arranged periodically with a period $p=400~\mathrm{nm}$. Nanostripes were $325~\mr{nm}$
wide, $(19\pm 2)~\mathrm{nm}$ thick and $80~\mathrm{\upmu m}$ long. A single stripe in the center of the MC was elongated on both sides by 8
$\mathrm{\upmu m}$ to increase its coercivity. By tuning the magnetic history, we magnetized the short stripes in $+y$-direction, while the
prolonged stripe was magnetized in $-y$-direction (Fig. \ref{fig:BLS_MC}(b)). In this state, the prolonged stripe is magnetized in opposite
direction compared to the rest of the MC, i.e., the short stripes, and we refer to it as \emph{magnetic defect}. On top of the MC two coplanar
waveguides (CPWs) with signal and ground line widths of $0.8~\mathrm{\upmu m}$ were prepared out of 5~nm thick Ti and 110~nm Au. For
phase-resolved spin wave transmission experiments based on all-electrical spectroscopy both CPWs were used. Spectra were reported in Ref.
\cite{Baumgaertl2018} (sample MC1). In the presented study we go beyond the earlier studies \cite{doi:10.1063/1.4953395,
Baumgaertl2018,doi:10.1021/acsami.9b02717} and exploit focused laser light to investigate microscopic aspects of the phase shifting process
with high spatial resolution. We excited SWs by applying a microwave current at CPW1 and used $\upmu\mr{BLS}$ for detection.
Figure \ref{fig:BLS_MC}(c) shows an optical image of MC1 taken with the $\upmu\mr{BLS}$ camera. We focused a laser with a wavelength of 473~nm
and 1~mW power to a spot with a diameter of about 350~nm onto the sample surface. The laser spot was scanned in $+x$-direction in 100~nm
steps, while SW intensity and phase were measured (see yellow scan path in Fig. \ref{fig:BLS_MC}(c)). A magnetic field $\upmu_0 H_\mr{B}$ was
applied in $+y$-direction, corresponding to the magnetization direction of the short stripes.\\
The 1D MC with magnetic defect was investigated for several $\upmu_0 H_\mr{B}$. Then the prolonged stripe was magnetized in $+y$-direction and
measurements were performed on the defect-free MC1 using identical instrument settings. The microwave frequency $f_\mr{ex}$ used for
excitation was adapted for each $H_\mr{B}$ in order to excite a SW with $k_1=2\uppi/\lambda=2.0~\mr{rad}\,\upmu\mr{m}^{-1}$
\cite{Baumgaertl2018}. Thereby we excite a SW with $\lambda = 3.1~\upmu\mr{m}$ which is more than 9 times larger than the width of the
magnetic defect. We use micromagnetic simulations with MuMax3 \cite{doi:10.1063/1.4899186} to explore $\Delta \Theta$ and $\eta$ for a wide
range of bias fields and wave vectors. We simulated a slice of MC1 in the $x$-$z$-plane, while in $y$-direction the periodic boundary
condition (PBC) approach \cite{doi:10.1063/1.3068637} with 1024 repetitions in $+y$ and $-y$-direction was applied, assuming a constant
magnetization of nanostripes along their lengths. We used $\upmu_0 M_\mr{s}=1.8~\mr{T}$ as saturation magnetization \cite{Baumgaertl2018},
$\alpha=0.006$ as Gilbert damping \cite{doi:10.1063/1.2337165}, $A_\mr{ex}=20~\mr{pJ\,m^{-1}}$ as exchange constant
\cite{doi:10.1063/1.4967826} and a grid size of $2~\mr{nm}\times 20~\mr{nm}\times 2~\mr{nm}~(\Delta x, \Delta y, \Delta z)$. For band
structure simulations a chain of 40 stripes and for SW propagation a chain of 164 stripes was considered.\\Following Refs.
\cite{doi:10.1063/1.3562519,Kumar2017} we simulated band structures by exciting the MC with a spatially and temporally varying magnetic field
(given by sinc functions) and subsequent computation of the Fourier amplitudes of the dynamic magnetization components $m_x(x,t)$ and
$m_y(x,t)$. We obtained a good agreement with the measured band structure of MC1 \cite{Baumgaertl2018}, when the simulated film thickness was
reduced to $d=10~\mr{nm}$ (supplement Fig. S1). The discrepancy with the nominal value of $d$ might be due to film roughness in the real
sample, which reduced the surface pinning \cite{Mercier1996} and was not considered in the simulations. \\
For simulating SW transmission through a magnetic defect, SWs were excited locally at $x=0~\mr{\upmu m}$ by a sinusoidal $h_\mr{rf}$ exploring
different frequencies $f_\mr{ex}$ at differnt $H_\mr{B}$. An individual stripe at $x_\mr{D}=6~\mr{\upmu m}$ was magnetized in $-y$-direction.
The other stripes were magnetized in $+y$-direction. We analyzed amplitudes and phases of $m_x(x,t)$ and $m_z(x,t)$ after the simulations had run for $t'$=10~ns and propagating SWs had reached a steady state. To avoid back-reflection, an absorbing boundary condition following Ref.
\cite{venkat2018absorbing} was applied at the outer edges of the MC.\\
Figure \ref{fig:BLS_MC}(d) displays the SW phase signal measured for MC1 with magnetic defect (green lines) and without (blue lines) at
specific $\upmu_0 H_\mr{B}$. Phase-resolved $\upmu\mr{BLS}$ allowed us to measure $\cos(\Theta(x)-\Theta')$, where $\Theta(x)$ is the SW phase
at a position $x$ and $\Theta'$ is a reference phase\cite{doi:10.1063/1.3115210}. $\Theta'$ is  constant for a given frequency $f_\mr{ex}$. For all
$\upmu_0 H_\mr{B}$, we observed sinusoidal waves with a wavelength $\lambda \simeq 3.1~\upmu\mr{m}$. In the defect-free state (all stripes
magnetized in one direction), the sinusoidal wave profile was unperturbed at the position $x_\mr{D}$ of the prolonged stripe. We assume that
due to the large aspect ratio of the investigated nanostripes, the demagnetization factor in $y$-direction was already close to zero for all
nanostripes \cite{doi:10.1063/1.367113} and the additional prolongation had little impact on the nanostripe's demagnetization field. When the
prolonged stripe was oppositely magnetized (magnetic defect state) the phase was clearly modified at $x=x_\mr{D}$. For $\upmu_0
H_\mr{B}=0~\mr{mT}$ (top row in Fig. \ref{fig:BLS_MC}(d)) a localized phase jump (dip) was observed at $x_\mr{D}$. For $x>x_\mr{D}$, the phase
profiles with and without defect were still in good agreement. We attribute the localized phase jump at $x_\mr{D}$ to an in-plane dynamic
coupling of the defect's magnetization to its neighboring stripes, as suggested in Huber \emph{et al.} \cite{Huber2013}. Due to magnetic
gyrotropy, the sense of spin-precessional motion in the defect is opposite to the rest of the MC. Consequently, the in-phase coupling results
in a $\uppi$ phase jump of the dynamic out-of-plane magnetization component, which is detected by $\upmu\mr{BLS}$.\\\begin{figure}[b!]
	\centering
	\includegraphics{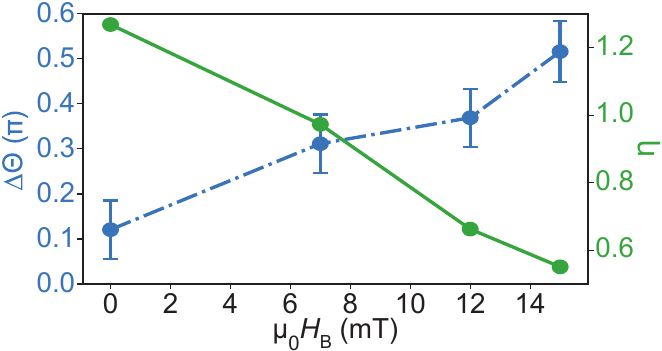}
	\caption{Measured phase shift $\Delta\Theta$ (blue line) and attenuation ratio  $\eta$ (green) for SWs at $x>x_\mr{D}$ with magnetic
defect compared to SWs without defect.}
	\label{fig:Fig2}
\end{figure}
For increasing $\upmu_0 H_\mr{B}$ (second and third row in Fig. \ref{fig:BLS_MC}(d)), the phase profiles with and without defect were
significantly displaced relative to one another for $x>x_\mr{D}$. The displacements indicate phase shifts of SWs. Strikingly, the relative
displacements were pronounced directly at the defect, i.e. the phase shifts were established on the length scale of the individual stripe of
width $w=325~\mr{nm}$. We quantify phase shifts $\Delta \Theta$ by fitting cosine functions $\cos(\Theta(x)-\Theta')$ for SWs which passed the
defect (yellow) and SWs in the defect-free state (black dashed lines in Fig. \ref{fig:BLS_MC}(d)). The blue line in Fig. \ref{fig:Fig2}
displays the extracted $\Delta \Theta$ as a function of $\upmu_0 H_\mr{B}$. In good agreement with Ref. \cite{Baumgaertl2018}, we find a
monotonous increase of $\Delta \Theta$ with $\upmu_0 H_\mr{B}$ reaching a phase shift close to $\uppi/2$ at 15~mT. The experiments of Fig.
\ref{fig:BLS_MC}(d) reveal the nanoscale nature of the phase shifting mechanism. The phase shift was concomitant with a reduction in the
amplitude of transmitted SWs (details in Supplement Fig. S2). The green line in Fig. \ref{fig:Fig2} displays the ratio $\eta$ of measured SW
amplitudes with and without defect for $x>x_\mr{D}$. At $15$~mT the defect reduced the SW amplitude by a factor of 2.\\
\begin{figure}[bt]
	\centering
	\includegraphics{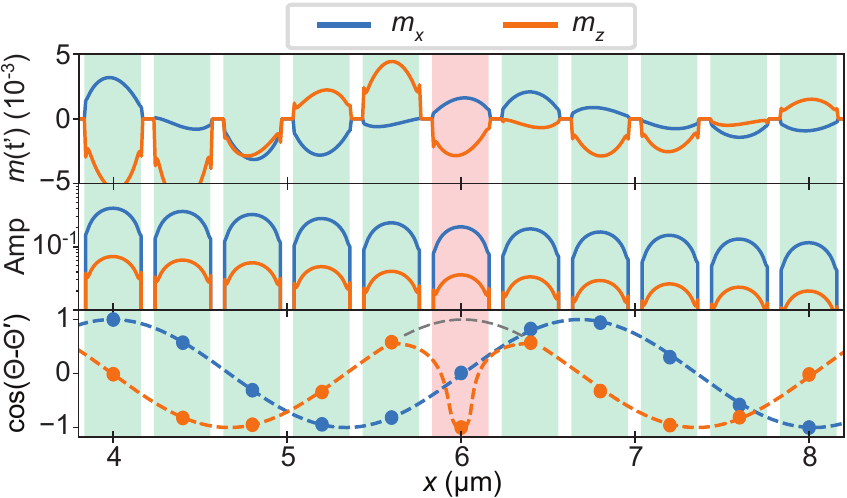}
	\caption{Simulated $m_x$ and $m_z$ for a propagating SW excited at $x= 0~\mr{\upmu m}$ with $k_x\approx 2~\mr{rad\,\upmu m^{-1}}$ shown
for $\upmu_0 H_\mr{B} = 0~\mr{mT}$. The stripe at $x_\mr{D}=6~\upmu\mr{m}$ (marked in red) was oppositely magnetized. The first row shows a
snapshot of $m_x$ and $m_z$ at $t'=10~\mr{ns}$. The second and third row depicts the precessional amplitudes and cosine of the phase. At the
defect a $\uppi$ phase jump of the phase of $m_z$ is observed.}
	\label{fig:Sim_Fig1}
\end{figure}
In Fig. \ref{fig:Sim_Fig1} we present micromagnetic simulations displaying results in the vicinity of the defect for $\upmu_0 H_\mr{B} =
0~\mr{mT}$ and a SW with $k_x\approx 2~\mr{rad\,\upmu m^{-1}}$. The first row in Fig. \ref{fig:Sim_Fig1} displays a snapshot of $m_x(x,t')$
and $m_z(x,t')$ at $t'$. Due to the ellipticity of the magnetization precession the amplitude of $m_z$ was small compared to $m_x$ and
multiplied by a factor of 10 for better visibility. We recorded $m_x$ and $m_z$ for $t\geq t'$ in 10~ps steps during a time span of 2~ns and
computed the FFT amplitude (Amp) and phase ($\Theta$) at the driving frequency $f_\mr{ex}$. Amp($m_x$) and Amp($m_z$) are displayed in the
second row of Fig. \ref{fig:Sim_Fig1}. The SWs decayed exponentially with a decay length $\delta=3.15~\upmu\mr{m}$, which was in good
agreement with $\delta=2.9\pm0.2~\upmu\mr{m}$ observed in the experiment (see Fig. S2). At the defect, no significant change in amplitude was
visible for $\upmu_0 H_\mr{B} = 0~\mr{mT}$. In the third row, we plot $\cos(\Theta-\Theta')$ for $\Theta(m_x)$ and $\Theta(m_z)$ extracted at
the center of each stripe. $\Theta(m_x)$ followed a sinusoidal wave with $\lambda\sim 3~\mr{\upmu m}$ (blue dashed line in Fig.
\ref{fig:Sim_Fig1}) without deviation at the defect. For $\Theta(m_z)$ however a local phase jump of $\uppi$ is apparent at the defect (orange
dashed line  in Fig. \ref{fig:Sim_Fig1}), which agrees well with the measurement observation on MC1 for $\upmu_0 H_\mr{B} = 0~\mr{mT}$ (cf.
first row in Fig. \ref{fig:BLS_MC}(d)).
\begin{figure*}[t!]
	\centering
	\includegraphics{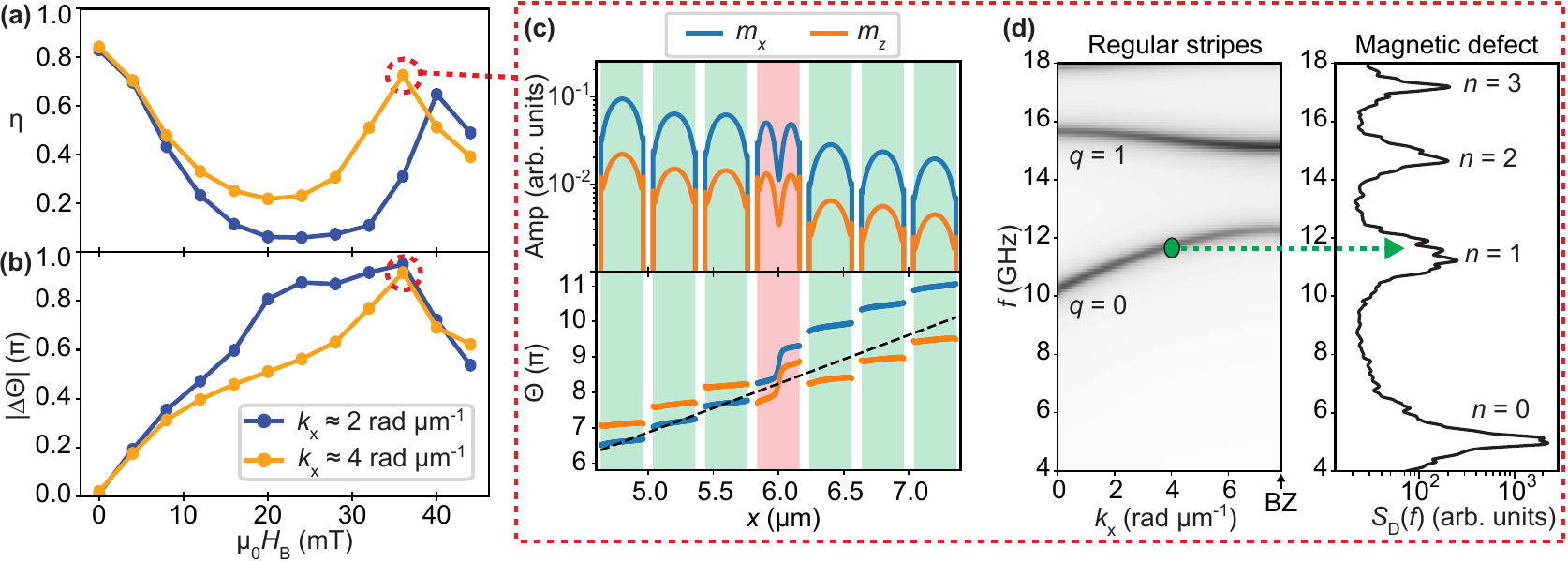}
	\caption{(a) Simulated amplitude ratio $\eta$ and (b) phase shift $|\Delta \Theta|$ as function of an applied bias field. For SWs with
$k_x \approx 4~\mr{rad}\, \upmu\mr{m}^{-1}$ a large $\eta$ and close to $\uppi$ phase shift was achieved at 36~mT (marked by red circle). In
(c) we display the respective amplitude and phase evolution. The dynamic magnetization profile at the defect indicates a laterally standing
mode with quantization number $n=1$. (d) Dispersion relation of the 1D MC and the power spectral density $S_\mr{D}(f)$ of thermally excited
magnons at the defect simulated for $\upmu_0 H_\mr{B} = 36~\mr{mT}$, respectively. The frequency of SWs with $4~\mr{rad}\, \upmu\mr{m}^{-1}$
in the first miniband ($q=0$) of the MC matches with the frequency of the second spin wave resonance $n=1$ of the defect.}
	\label{fig:Sim_Fig2}
\end{figure*}
\\For a quantitative analysis of amplitude and phase changes introduced by the defect, we repeated simulations for the defect-free state as a
reference. We computed $\eta$ as the ratio of Amp($m_x$) values with and without defect. Further we computed the magnitude of the phase shift
$|\Delta\Theta|$ based on the difference of $\Theta(m_x)$ with and without defect. Both $\eta$ and $|\Delta\Theta|$ were evaluated in the
region $x=8$ to $12~\upmu\mr{m}$ and then averaged. Bias fields from 0 to 44~mT were simulated in 4~mT steps. We note that in the simulations
the defect was not switched up to 44~mT, while in our experiments the defect switched at 23~mT. Switching fields in real stripes with rough
edges have already been reported to be smaller compared to stripes with ideal edges in simulations \cite{Topp2011}.
For each $H_\mr{B}$, we computed the dispersion relation and extracted frequencies of the first miniband at $k_x = k_1= 2~\mr{rad}\,
\upmu\mr{m}^{-1}$ (as used in the experiment) and $k_x = 4~\mr{rad}\, \upmu\mr{m}^{-1}$ (in the middle of the first Brillouin zone of the MC).
Then SW propagation was simulated for the extracted frequencies. In this manner, we evaluated $\eta$ and  $|\Delta\Theta|$ as a function of
$H_\mr{B}$ without significantly varying the wave vector (Fig. \ref{fig:Sim_Fig2}(a) and (b)). For $k_x \approx 2~\mr{rad} \,\upmu\mr{m}^{-1}$
we observe a decrease in transmission with $H_\mr{B}$ until $\upmu_0 H_\mr{B} = 24~\mr{mT}$, where $\eta$ reaches a minimum value of 0.06. In
the same field regime we extract an approximately linear increase of $|\Delta\Theta|$ from 0 to $0.88\uppi$, which is in good qualitative
agreement with our experimental data (Fig. \ref{fig:Fig2}).\\ Strikingly, above 24~mT the simulated transmission coefficient $\eta$ started to
increase with $H_\mr{B}$. Concomitantly $|\Delta\Theta|$ further increased. For $k_x = 2~\mr{rad}\, \upmu\mr{m}^{-1}$ we found $\eta=0.65$ at
40~mT. $|\Delta\Theta|$ peaked at 36~mT, amounting to $0.95\uppi$. For $k_x \approx 4~\mr{rad} \,\upmu\mr{m}^{-1}$, the maximum in
$|\Delta\Theta|$ coincided with the local maximum in $\eta$ for $\upmu_0 H_\mr{B} = 36~\mr{mT}$. We found $|\Delta\Theta|=0.92\uppi$ and an
appreciable transmission with $\eta=0.73$, allowing for low-loss phase inversion.\\In the following we discuss the origin of the large $\eta$
for high $H_\mr{B}$. Figure \ref{fig:Sim_Fig2}(c) shows Amp and $\Theta$ for SWs with $k_x \approx 4~\mr{rad} \,\upmu\mr{m}^{-1}$ excited at
$f_\mr{ex} = 11.73~\mathrm{GHz}$ and $\upmu_0 H_\mr{B} = 36~\mr{mT}$. The plotted $\Theta$ has been unwrapped and the slope (black dashed
line) represents $\Theta-\Theta'=k_xx$. The phase evolution of stripes neighboring the defect behaves regularly. The phase shift occurs right
at the center of the defect, where $\Theta$ abruptly shifts by about $\uppi$. At the same position, a node in the SW amplitude is observed.
The dynamic magnetization profile along the width of the defect agrees well with a standing wave with quantization number $n=1$
\cite{BoundaryCondition}. To identify the eigenfrequencies of the magnetic defect, we simulated with MuMax3 thermally excited magnons at a
finite temperature $T=300~\mr{K}$ \cite{doi:10.1063/1.5003957}. The simulation was run over an extended time period of 100~ns and then the
power spectral density $S_\mr{D}(f)$ of $m_x(t)$ at the position of the defect was computed. Allowed SW eigenfrequencies are apparent as peaks
in $S_\mr{D}(f)$ \cite{doi:10.1063/1.5003957, doi:10.1063/1.1402146, ThermalNoise}. By considering thermal magnons we are not limited to SW
modes compatible with the symmetry of an exciting $h_{\mr{rf}}$. \\ Figure \ref{fig:Sim_Fig2}(d) compares the band structure of the 1D MC and
$S_\mr{D}(f)$ of the magnetic defect for $\upmu_0 H_\mr{B} = 36~\mr{mT}$. Here the frequency of SWs with $k_x=4~\mr{rad} \,\upmu\mr{m}^{-1}$
in the first miniband ($q=0$) matches well with the frequency of the second allowed state ($n=1$) observed in $S_\mr{D}(f)$. In the contrary
for 24~mT, where the transmission was low, the relevant  $f_\mr{ex}$  was between the $n=0$ and $n=1$ peaks of $S_\mr{D}(f)$ (Supplement Fig.
S3), i.e. inside a forbidden frequency gap. Our finding suggests that high transmission is obtained when one of the eigenfrequencies of the
defect is resonantly tuned to $f_\mr{ex}$. We speculate that for lager $H_\mr{B}$ not considered in the simulation, further maxima in $\eta$
are achieved every time the frequency of SWs excited in a miniband $q$ of the regular magnetized stripes overlaps with a higher eigenfrequency
state $n=q+i$ (with $i\in\mathbb{N}$ and $i\geq2$) at the defect. Based on the dynamic magnetization profiles know for laterally standing
waves in a nanostripe \cite{PhysRevLett.81.SingleStripBLS,BoundaryCondition,Gubbiotti2005}, we anticipate a phase shift of $\sim\uppi$ in case
$n - q$ is odd and $\sim 0\uppi$ in case $n - q$ is even.\\
To conclude, via phase-resolved $\upmu$BLS we measured the phase evolution of propagating SWs in a 1D MC consisting of dipolarly-coupled
nanostripes. When a single nanostripe was magnetized in opposing direction, a local phase jump of the out-of-plane dynamic component was
detected. For $\upmu_0 H_\mr{B}> 0~\mr{mT}$, phase shifts occurred on the length scale of 325~nm much smaller than $\lambda$, and were
concomitant with a reduction in transmission amplitude. Using micromagnetic simulations we found however an increase in transmission, once the
bias field was sufficient to align the magnon miniband with the eigenfrequency of the second laterally quantized mode in the defect. Due to
the resonant coupling of the defect, a high transmission and a phase shift of close to $\uppi$ were achieved, allowing for a low-loss phase
inverter. For future experimental studies, it will be relevant to either increase the switching field (e.g. by using a material with an
appropriate magnetocrystalline anisotropy) or to reduce the frequency spacing of the minibands. The latter could be realized for 1D MCs
prepared from Yttrium iron garnet. Our results pave the way for efficient and low-loss phase inverters in nanomagnonics.\\\\
\textbf{Supplementary Material}\\
See supplementary material for a comparison of the simulated and experimental dispersion relation, SW intensities measured with $\upmu$BLS,
and the simulated dispersion relation and $S_\mr{D}(f)$ at $\upmu_0 H_\mr{B} = 24~\mr{mT}$.  \\\\
\textbf{Acknowledgment}\\ We thank for funding by SNSF via grant 163016. We thank F. Stellacci, E. Athanasopoulou and S. Watanabe for support
concerning MFM.\\\\
\textbf{Data Availability}\\
The data that support the findings of this study are openly available in Zenodo at \href{http://doi.org/10.5281/zenodo.4680409}{http://doi.org/10.5281/zenodo.4680409}.
%
\end{document}